\documentclass[]{jpsj2}

\title{Convergence of Quantum Annealing with Real-Time
Schr\"{o}dinger Dynamics}
\author{Satoshi \textsc{Morita} and Hidetoshi \textsc{Nishimori}}

\inst{Department of Physics, Tokyo Institute of Technology, Oh-okayama,
Meguro-ku, Tokyo 152-8551}

\abst{Convergence conditions for quantum annealing are derived for
  optimization problems represented by the Ising model of a general
  form.  Quantum fluctuations are introduced as a transverse field
  and/or transverse ferromagnetic interactions, and the time evolution
  follows the real-time Schr\"{o}dinger equation.  It is shown that the
  system stays arbitrarily close to the instantaneous ground state,
  finally reaching the target optimal state, if the strength of quantum
  fluctuations decreases sufficiently slowly, in particular inversely
  proportionally to the power of time in the asymptotic region.  This is
  the same condition as the other implementations of quantum annealing,
  quantum Monte Carlo and Green's function Monte Carlo simulations, in
  spite of the essential difference in the type of dynamics.  The method
  of analysis is an application of the adiabatic theorem in conjunction
  with an estimate of a lower bound of the energy gap based on the
  recently proposed idea of Somma {\it et. al.} for the analysis of
  classical simulated annealing using a classical-quantum
  correspondence.  }

\kword{quantum annealing, annealing schedule,
adiabatic theorem, optimization problem, transverse-field Ising model}

\begin{document}
\maketitle

Quantum annealing (QA) recently attracts much attention as a novel
algorithm for optimization problems. \cite{KN, DC, ST, FGSSD} A
fictitious kinetic energy of quantum nature is introduced to the
classical system which represents the cost function to be minimized. The
resulting system searches the phase space by means of quantum
transitions, which are gradually decreased as time proceeds. If the
initial state is the ground state of the initial quantum Hamiltonian,
the system is expected to keep track of the ground state of the
instantaneous Hamiltonian under a slow decrease of quantum fluctuations.
From this viewpoint, QA is also called quantum adiabatic evolution.
\cite{FGGS} Most of the numerical studies \cite{KN, DC, ST, SMTC, MST,
SO, SPA, LB, LB2, MST2, SST05, SST06, DCS} showed that QA is more
efficient in solving optimization problems than the well-known classical
algorithm, simulated annealing (SA). \cite{KGV,AK}

Convergence theorems for stochastic implementations of QA have been
proved for the transverse-field Ising model. \cite{MN} A power-law
decrease of the transverse field has been shown to be sufficient to
guarantee convergence to the optimal state for generic optimization
problems.  This power-law annealing schedule is faster than that of the
inverse-log law for SA given in the theorem of Geman and
Geman. \cite{AK, GG} However, these theorems for QA were proved for
stochastic processes to realize QA.  It has been unknown so far what
annealing schedule would guarantee the convergence of QA following the
real-time Schr\"{o}dinger equation.  We have solved this problem on the
basis of the idea of Somma {\it et. al.}  \cite{SBO} These authors found
that the inverse-log law condition for SA can be derived from the
adiabatic theorem for a quantum system obtained from the original
classical system through a classical-quantum mapping.  Although they
also discussed some aspects of QA, their interest was to use quantum
mechanics to simulate finite-temperature classical statistical
mechanics.  We point out in the present article that the convergence
condition of genuine quantum annealing, in which the system follows the
real-time Schr\"odinger equation without temperature, can also be
derived by a similar analysis.

Let us suppose that the optimization problem one wants to solve can be
represented as the ground-state search of an Ising model of general form
\begin{equation}
  H_{\rm pot}\equiv -\sum_{i=1}^N J_i\sigma_i^z
  -\sum_{ij}J_{ij}\sigma_i^z\sigma_j^z
  -\sum_{ijk}J_{ijk}\sigma_i^z\sigma_j^z\sigma_k^z-\cdots,
  \label{eq:potential}
\end{equation}
where $\sigma_i^z$ denotes the $z$ component of the Pauli matrix at site
$i$.  Quantum annealing is realized typically by the addition of a
time-dependent transverse field
\begin{equation}
  H_{\rm kin}(t)\equiv -\Gamma(t) \sum_{i=1}^{N}\sigma_i^x,
  \label{eq:kinetic}
\end{equation}
which may be regarded as the quantum kinetic energy to be compared with
the potential energy (\ref{eq:potential}).
Initially the coefficient of the kinetic term $\Gamma (t)$ is chosen to be
very large, and the total Hamiltonian
\begin{equation}
  H(t)=H_{\rm pot}+H_{\rm kin}(t)
  \label{eq:total_H}
\end{equation}
is dominated by the second kinetic term.  The coefficient $\Gamma (t)$
is then decreased gradually toward 0, leaving eventually only the
potential term.  Accordingly the state vector $|\psi(t)\rangle$, which
follows the real-time Schr\"{o}dinger equation
\begin{equation}
  {\rm i}\frac{\rm d}{{\rm d}t}|\psi(t)\rangle=H(t)|\psi(t)\rangle,
\end{equation}
is expected to evolve from the trivial initial ground state of the
transverse field (\ref{eq:kinetic}) to finally the non-trivial ground
state of eq. (\ref{eq:potential}).  The problem of central concern in
the present paper is how slowly we should decrease $\Gamma (t)$ to keep
the state vector arbitrarily close to the instantaneous ground state of
the total Hamiltonian (\ref{eq:total_H}), namely the adiabaticity
condition, to achieve the goal of minimization of
eq. (\ref{eq:potential}).

The adiabatic theorem \cite{Mess} provides the excitation probability at
time $t$ as
\begin{equation}
  \left|\langle n(t) | \psi(t) \rangle\right|^2 \simeq \left|
\frac{\left\langle n(t)\left| {\displaystyle
\frac{\partial H(t)}{\partial t}} \right|0(t)\right\rangle }
  {\left(\varepsilon_n(t)-\varepsilon_0(t)\right)^2}\right|^2,
  \label{eq:adiabatic}
\end{equation}
where $|n(t)\rangle$ is the $n$th instantaneous eigenstate of $H(t)$
with the eigenvalue $\varepsilon_n(t)$. We assume that $|0(t)\rangle$ is
the ground state of $H(t)$ and $|\psi(0)\rangle=|0(0)\rangle$. The
probability (\ref{eq:adiabatic}) needs to be arbitrarily small for the
success of QA.  We therefore evaluate an upper bound of
\begin{equation}
   \frac{\left| \left\langle n(t)\left|
  \displaystyle\frac{\partial H(t)}{\partial t}
  \right| 0(t)\right\rangle \right|}
  {\left(\varepsilon_n(t)-\varepsilon_0(t)\right)^2} .
  \label{eq:condition}
\end{equation}

For this purpose we estimate the numerator and the denominator of
eq. (\ref{eq:condition}).  As for the numerator, it is straightforward
to see
\begin{equation}
  \left| \left\langle n(t)\left|\frac{\partial H(t)}{\partial t}
  \right| 0(t)\right\rangle \right| \leq -N \frac{{\rm d}\Gamma}{{\rm d}t},
  \label{eq:numerator}
\end{equation}
since the time dependence of $H(t)$ lies only in the kinetic term,
which has $N$ terms.
Note that ${\rm d}\Gamma /{\rm d}t$ is negative.

A lower bound on the denominator of eq. (\ref{eq:condition}) can be
evaluated using an inequality for a strictly positive
operator.\cite{Hopf} First we recall that the Perron-Frobenius theorem
states that a non-negative square matrix $M$ has a real eigenvalue
$\lambda_0$ satisfying $|\lambda|\le\lambda_0$ for any other eigenvalue
$\lambda$.  If all the elements of $M$ are strictly positive,
$M_{ij}>0$, its eigenvalues satisfy the stronger inequality \cite{Hopf},
\begin{equation}
  |\lambda|\leq\frac{\kappa-1}{\kappa+1}\lambda_0,
   \label{eq:Hopf}
\end{equation}
where $\kappa$ is defined by
\begin{equation}
  \kappa= \max_{i,j,k}\frac{M_{ik}}{M_{jk}}.
\end{equation}
We apply the above inequality (\ref{eq:Hopf}) to the operator $M\equiv
(E_{\rm max}-H(t))^N$, where $E_{\rm max}$ is the largest eigenvalue of
$H_{\rm pot}$.  All the elements of the matrix $M$ are strictly positive
in the representation that diagonalizes $\sigma_i^z$ because $E_{\rm
max}-H(t)$ is non-negative and irreducible (that is, any state can be
reached from any other state within $N$ steps at most).  In the
asymptotic region $t\gg 1$ where $\Gamma(t)\ll 1$, the minimum element
of $M$, which is between two states having all spins in mutually
opposite directions, is equal to $N!  \Gamma(t)^N$, where $N!$ comes
from the ways of permutation to flip spins.  Replacement of $H_{\rm
kin}$ by $-N$ shows that the maximum matrix element of $M$ has the upper
bound $(E_{\rm max}-E_{\rm min}+N)^N$, where $E_{\rm min}$ is the lowest
eigenvalue of $H_{\rm pot}$.  Thus we have
\begin{equation}
  \kappa\leq\frac{(E_{\rm max}-E_{\rm min}+N)^N}{N! \Gamma(t)^N}.
  \label{eq:kappa}
\end{equation}
Since the eigenvalue of $H(t)$ is denoted by $\varepsilon_n(t)$,
eq. (\ref{eq:Hopf}) is rewritten as
\begin{equation}
  (E_{\rm max}-\varepsilon_n(t))^N\leq \frac{\kappa-1}{\kappa+1}
  (E_{\rm max}-\varepsilon_0(t))^N.
\end{equation}
Substitution of eq. (\ref{eq:kappa}) into the above inequality yields
\begin{equation}
  \varepsilon_n(t)-\varepsilon_0(t) \ge
  \frac{2 (E_{\rm max}-\varepsilon_0(t)) N! }
  {N (E_{\rm max}-E_{\rm min}+N)^N} \Gamma(t)^N \equiv A\Gamma(t)^N,
  \label{eq:gap}
\end{equation}
where we used $\kappa\gg 1$ in the asymptotic time region where 
$\Gamma (t)$ is very small.
The coefficient $A$ is estimated using the Stirling formula as
\begin{equation}
  A\simeq \frac{2\sqrt{2\pi N} (E_{\rm max}-\varepsilon_0(t))}{N {\rm e}^N}
  \left(\frac{N}{E_{\rm max}-E_{\rm min}+N}\right)^N,
\end{equation}
which demonstrates that $A$ is exponentially small for large $N$.

Now, by the combination of the above estimates (\ref{eq:numerator}) and
(\ref{eq:gap}), we find that the sufficient condition for convergence is that
the upper bound of eq. (\ref{eq:condition})
\begin{equation}
  -\frac{N}{A^2 \Gamma(t)^{2N}}\frac{{\rm d}\Gamma}{{\rm d}t}
  \label{eq:upperbound}
\end{equation}
is arbitrarily small.
By equating eq. (\ref{eq:upperbound}) to a small constant $\delta$ and
integrating the resulting differential equation, we find
\begin{equation}
  \Gamma(t) = (\alpha\, t)^{-1/(2N-1)},
   \label{eq:AS}
\end{equation}
where $\alpha$ is exponentially small for large $N$ and is proportional
to $\delta$.  The transverse field should be decreased following this
functional form or slower.  Therefore the asymptotic power decay of the
transverse field guarantees that the excitation probability is bounded
by the arbitrarily small constant $\delta^2$ at each time.

The same discussions apply to quantum annealing using transverse
ferromagnetic interactions in addition to a transverse field,
\begin{equation}
  \tilde{H}_{\rm kin}(t)=-\tilde{\Gamma}(t)
   \left(\sum_{i=1}^N\sigma_i^x
    +\sum_{ij}\sigma_i^x \sigma_j^x\right).
  \label{eq:TI}
\end{equation}
A recent study showed the effectiveness of this type of quantum kinetic
energy \cite{Suzuki}.  A modification of the strictly positive operator
to $(E_{\rm max}-H(t))^{N/2}$ in the above argument leads to a lower
bound of the energy gap as a quantity proportional to
$\tilde{\Gamma}(t)^{N/2}$.  The resulting asymptotic annealing schedule
is
\begin{equation}
  \tilde{\Gamma}(t)\propto t^{-1/(N-1)},
\end{equation}
which is faster than the case with the transverse field only,
eq. (\ref{eq:AS}).  This result implies that the additional non-zero
off-diagonal elements of $H(t)$ would widen the energy gap and
accelerate the convergence of QA.
\\

In this paper, we have derived conditions for convergence of QA under
the real-time Schr\"{o}dinger dynamics using the adiabatic theorem. The
asymptotic power-law annealing schedule (\ref{eq:AS}) guarantees the
adiabatic evolution during the annealing process at its final stage
$t\gg 1$.  This condition coincides with our previous results for
stochastic implementations of QA \cite{MN}.  It is remarkable that
essentially different types of dynamics share the same condition for
convergence.  Note that the power decay derived above applies to the
asymptotic region $t\gg 1$.  At the initial stage, $\Gamma (t)$ must be
tuned following a different functional form to satisfy the adiabatic
condition.

For the adiabatic theorem to be applicable, the energy gap between the
ground state and the first excitation state should be finite.  The
inequality (\ref{eq:gap}) implies that this condition is always
satisfied in the transverse-field Ising model as long as the system size
is finite. In the thermodynamic limit, of course, the gap may vanish at
the critical point.  We emphasize that the system size $N$ is kept
finite in the present paper because our purpose is to study optimization
problems in which the number of elements is always finite.

We thank Prof. G. E. Santoro for useful comments and discussions.  This
work was partially supported by CREST, JST and by the Grant-in-Aid for
Scientific Research on Priority Area `Deeping and Expansion of
Statistical Mechanical Informatics' by the Ministry of Education,
Culture, Sports, Science and Technology. One of the authors (SM) is
supported by JSPS Research Fellowships for Young Scientists.

\end{document}